\documentclass{article}
\usepackage{spconf,amsmath,graphicx}

\usepackage[dvipsnames]{xcolor}
\usepackage[pdf]{pstricks}
\usepackage{booktabs}
\usepackage{cleveref}
\usepackage{tikz-qtree}
\usepackage{multirow}
\usepackage{caption}
\usepackage{subcaption}
\usepackage{array}
\usepackage[textwidth=17mm]{todonotes}
\definecolor{red}{rgb}{0.8500, 0.3250, 0.0980}
\definecolor{blue}{rgb}{0.000, 0.4470, 0.7410}

\title{Distribution augmentation for low-resource expressive text-to-speech}

\name{Mateusz Lajszczak, Animesh Prasad, Arent van Korlaar, Bajibabu Bollepalli, Antonio Bonafonte}
\nameplus{Arnaud Joly, Marco Nicolis, Alexis Moinet, Thomas Drugman, Trevor Wood, Elena Sokolova}
\address{Alexa AI, Amazon, Cambridge}

\begin{document}
\ninept
\maketitle
\begin{abstract}

This paper presents a novel data augmentation technique for text-to-speech (TTS), that allows to generate new (text, audio) training examples without requiring any additional data.
Our goal is to increase diversity of text conditionings available during training.
This helps to reduce overfitting, especially in low-resource settings.
Our method relies on substituting text and audio fragments in a way that preserves syntactical correctness.
We take additional measures to ensure that synthesized speech does not contain artifacts caused by combining inconsistent audio samples.
The perceptual evaluations show that our method improves speech quality over a number of datasets, speakers, and TTS architectures.
We also demonstrate that it greatly improves robustness of attention-based TTS models.

\end{abstract}


\section{Introduction}
\label{sec:intro}

Data augmentation techniques play an important role in the training of neural networks, particularly in classification tasks \cite{cubuk2020randaugment,perez2017effectiveness, shorten2019survey}.
They allow to improve performance of the data-hungry models and reduce overfitting. 
Despite a rapid progress of neural technologies in the field of text-to-speech (TTS) \cite{oord2016wavenet, oord2018parallel, taco2}, very little work is devoted to applying data augmentation in this domain. 
At the same time, in many cases TTS data is scarce (e.g. for low-resource languages), thus data augmentation has the potential to be particularly impactful in such scenarios.

Data augmentation is underexplored in TTS partly because TTS is a generative problem. As argued in \cite{jun2020distribution}, transformations used for augmenting training data (e.g. in computer vision domain) often modify its distribution significantly. Thus, using them to train generative models without any countermeasures might lead to fitting the wrong distribution and, in result, to producing out-of-distribution samples during inference. Unsurprisingly then, data augmentation approaches for TTS focus mostly on generating more in-distribution data by applying techniques like voice conversion (VC) \cite{huybrechts2021low, shah21_ssw}.
Recent work on data augmentation for generative modelling attempts to overcome aforementioned limitations. In \cite{jun2020distribution}, a new method called distribution augmentation (DistAug) is proposed, where the generative model is conditioned on the augmentation itself. Thus, instead of a single distribution $p(x)$, a conditional distribution is learned $p(t(x) \mid t)$, where $t$ represents a transformation used to augment data. Provided that the identity transformation, $\mathrm{id}_D$, is used during training, one can recover the model for unaugmented data by taking $p(x \mid \mathrm{id}_D)$.
As explained in \cite{jun2020distribution}, DistAug method can be interpreted as a form of data-dependent regularization that reduces overfitting. Moreover, one can see it as a form multi-task learning that improves generalization by introducing beneficial inductive biases.

We take inspiration from this method and propose augmenting a TTS dataset with new samples that can potentially be out-of-distribution. We consider a family of augmentations which generate new training examples by assembling new texts and audios from existing ones. More precisely, similarly as in \cite{subsub2021}, we extract parse trees of text conditionings and create new texts by applying subtree substitution. We also apply a corresponding substitution to audio features taking advantage of available text-audio alignment. While our method does not improve acoustic diversity of the data, it increases diversity of text conditionings.

Our main contributions are as follows: 1) We introduce a novel data augmentation technique for TTS that can be used to considerably increase the diversity of text conditionings that are presented to the model. To our knowledge, this is the first application of distribution augmentation technique in neural TTS; 2) We demonstrate that our method reduces model’s overfitting to input text and greatly improves robustness of attention-based sequence-to-sequence TTS models; 3) We perform perceptual evaluations showing that our method improves quality of generated speech. We demonstrate this improvement on different data settings (varying speakers and dataset sizes) and two model architectures (attention based sequence-to-sequence model and a model using externally provided durations).

\section{Related work} \label{sec:relwork}

One of the most popular classes of methods for improving TTS model performance in low-data settings is transfer learning (TL) \cite{chung2019semi,tu2019end}.
In TL, the main idea is to transfer the knowledge learned in the source task to the target task.

NTTS systems based on TL methods first develop multi-speaker models as a source task.
Then, the multi-speaker models are adapted to the target voice as a target task. TL methods improve the synthesis quality of low-resource target voices \cite{chung2019semi,latorre2019effect}.
However, they depend on the availability of multi-speaker datasets and complicate the training, as they often require fine-tuning with target speaker data.

To increase the amount of target speaker data, data augmentation approaches have been proposed \cite{shah21_ssw,huybrechts2021low}.
They are based on voice conversion (VC) models that allow to generate additional target speaker data provided that appropriate source speaker data is available.

The authors in \cite{huybrechts2021low, shah21_ssw} applied the CopyCat VC model \cite{karlapati20_interspeech} to generate expressive synthetic speech in low-resource data settings. A drawback of this approach is that it requires additional speech datasets (including multi-speaker datasets) to train VC models and to provide source data for conversion. Moreover, performance of this approach depends on performance of the VC model.

In contrast to aforementioned methods, the data augmentation approach proposed in this work does not depend on collecting additional speech data or training additional models (like VC models).
This can be especially beneficial for low-resource languages for which large multi-speaker datasets are not available.
Furthermore, our method can in principle be applied on top of the existing approaches.

Expressive speech datasets are particularly challenging for TTS models. 
Many approaches were recently proposed to improve modelling of expressive speech \cite{stanton2018predicting, aggarwal2020using, valle2020flowtron, neekhara2021expressive}.
Our approach focuses more on alleviating issues with modelling expressive speech in low-resource setting.
This includes issues like lack of robustness (in attention-based models) and lower signal quality that are particularly severe when amount of expressive data is limited.

\section{Proposed method}
\label{sec:majhead}

\subsection{Data augmentation through word permutations} \label{sec:perm}

In this section we describe a general case of augmenting a TTS dataset by permuting words from this dataset. We point out potential issues with this kind of augmentations and propose techniques to mitigate them.

In TTS, the task is to model audio features $X$ (such as mel-spectrograms) for the given text representation $c$ (for example, a sequence of characters or phonemes). We can cast this as a probabilistic modelling problem, where the goal is to estimate the conditional distribution $p(X \mid c)$. Tacotron-like autoregressive models further factorize this distribution into $\prod p(x_i \mid x_{<i},c)$. Given a dataset $D = \{(X_n, c_n)\}$ of independent data samples we can use standard methods like maximum likelihood estimation to fit $p(X \mid c)$.

Let us now consider the problem of augmenting $D$ with new samples that are derived from elements of $D$ by decomposing them into pieces (e.g. single words) and reassembling into new combinations. In this way we can generate practically unlimited number of new training examples $(X, c)$ in order to improve model performance and reduce overfitting. However, this poses a problem of generating out-of-distribution samples which would prevent us from fitting the right distribution. This can happen for the following reasons:

\begin{enumerate}
\item[(1)]\vspace{-0.1cm} The marginal distribution of text conditionings $c$ is modified by augmentation (augmented samples can be syntactically or semantically unsound).
\item[(2)]\vspace{-0.1cm} Assembled audio features have out-of-distribution local structure (audio joints, locations where audio pieces are joined, might sound unnatural).
\item[(3)]\vspace{-0.1cm} Assembled audio features have out-of-distribution global structure (overall prosody of assembled audio is inconsistent).
\end{enumerate}

In order to address (1) we only need to make sure that augmented dataset contains enough samples that follow the original distribution of $c$ and the model has enough capacity. As argued in \cite{jun2020distribution}, samples with out-of-distribution conditionings should not prevent the model from properly fitting in-distribution samples.

In order to address (1) and (3) we consider the following form of data augmentation.
When sampling augmented data, we impose constraints on a distribution of augmented conditioning so that it does not diverge much from the original distribution of conditioning (i.e. often enough it follows rules of syntax and grammar).
Additionally we ensure that augmented examples are assembled from long enough text/audio fragments so our models can learn consistent prosody.
To implement augmentation following these constraints, we take inspiration from the substructure substitution method \cite{subsub2021} and propose to extend it for TTS.
We elaborate on it more in \Cref{sec:parse}.

To counter (2) and (3), we introduce an additional conditioning that identifies the type of augmentation.
This was originally proposed in \cite{jun2020distribution} as a global conditioning on augmentation transformation identity.
Since the global conditioning does not provide any information about the location of audio joints, we instead use a localized conditioning on audio joints, as depicted in \Cref{architecture}.
It identifies the type of augmentation and at the same time addresses (2).

\subsection{Constituency parse based tree substitutions}
\label{sec:parse}

As mentioned in the previous section, we look for a constraint on the text generation process that would ensure that augmented text is often enough syntactically and grammatically correct. Below we describe the text generation technique of our choice that satisfies this requirement.

\begin{figure}
    \centering
    \begin{subfigure}[t]{0.32\linewidth}
        \centering
        \begin{tikzpicture}[scale=0.5]
            \Tree[.S    [.NP [.PRP \textcolor{red}{He} ]]
           [.ADVP [.RB \textcolor{red}{never} ]]
           [.\textbf{\textbf{VP}} [.\textbf{VBD} \textcolor{red}{\textbf{lied}} ]]
            ]
        \end{tikzpicture}
        \includegraphics[width=14mm, height=10mm]{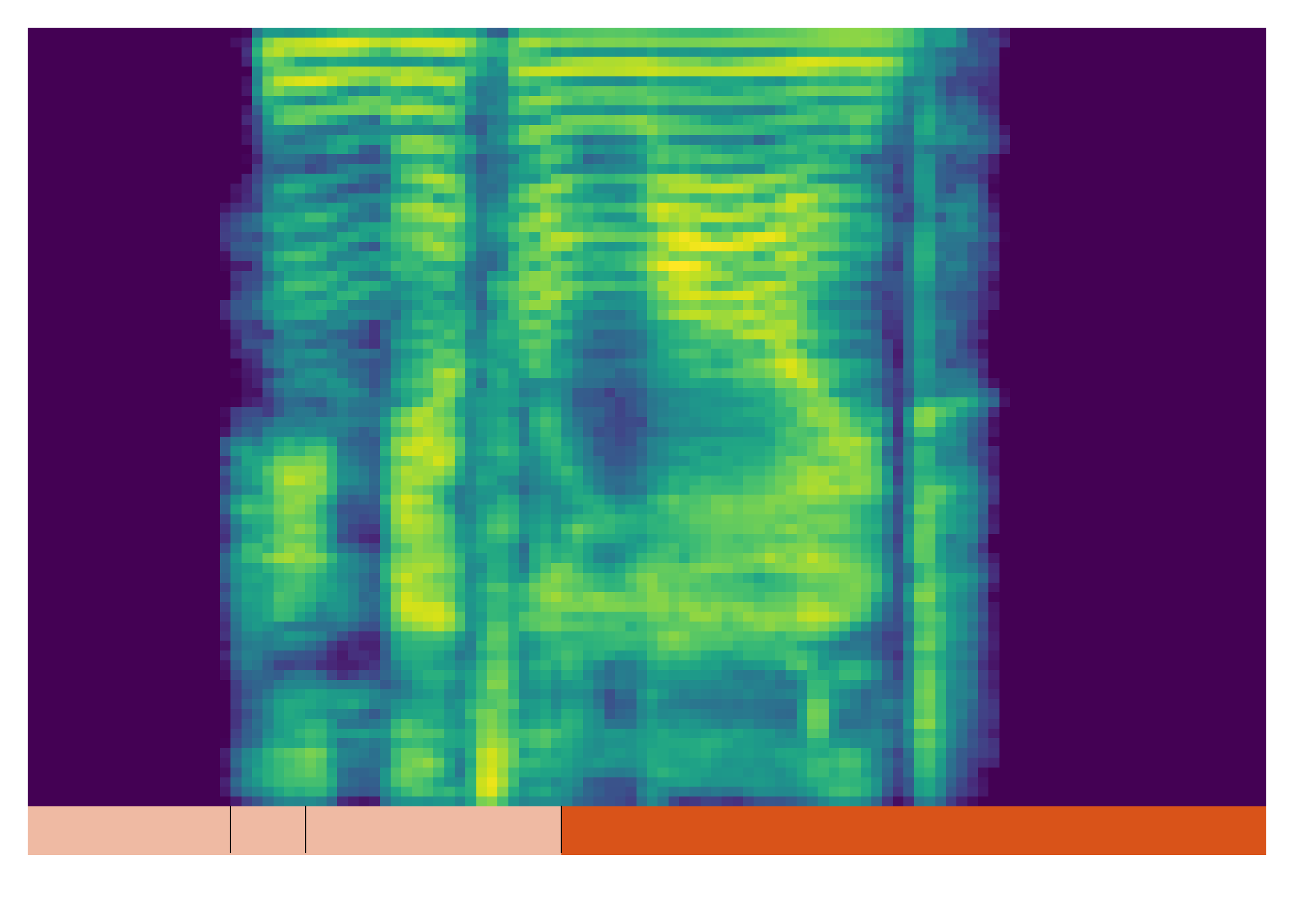}
        \caption{Utterance 1}
    \end{subfigure}
    \begin{subfigure}[t]{0.32\linewidth}
        \centering
        \begin{tikzpicture}[scale=0.5]
            \Tree[.S    [.NP [.PRP \textcolor{blue}{She} ]]
                    [.\textbf{VP}    [.\textbf{VBD} \textbf{\textcolor{blue}{shook}} ]
                            [.\textbf{NP} [.\textbf{PRP\$} \textbf{\textcolor{blue}{her}} ]
                                    [.\textbf{NN} \textbf{\textcolor{blue}{head}} ]]]
            ]
        \end{tikzpicture}
	\includegraphics[width=21mm, height=10mm]{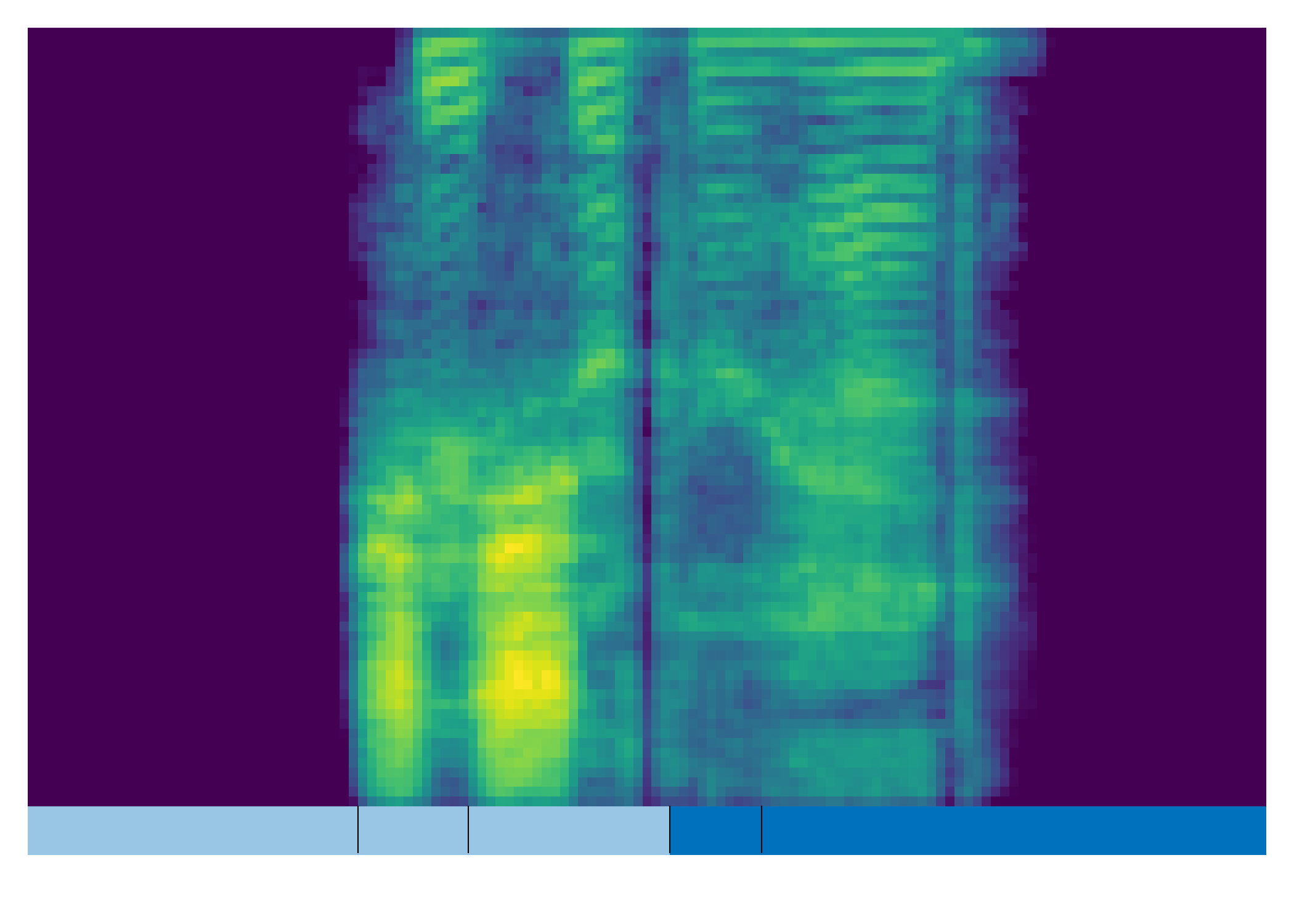}
        \caption{Utterance 2}
    \end{subfigure}
    \begin{subfigure}[t]{0.32\linewidth}
        \centering
        \begin{tikzpicture}[scale=0.5]
            \Tree[.S    [.NP [.PRP \textcolor{red}{He} ]]
                       [.ADVP [.RB \textcolor{red}{never} ]]
                       [.\textbf{VP}    [.\textbf{VBD} \textbf{\textcolor{blue}{shook}} ]
                               [.\textbf{NP} [.\textbf{PRP\$} \textbf{\textcolor{blue}{her}} ]
                                       [.\textbf{NN} \textbf{\textcolor{blue}{head}} ]]]
               ]
        \end{tikzpicture}
        \includegraphics[width=27mm, height=10mm]{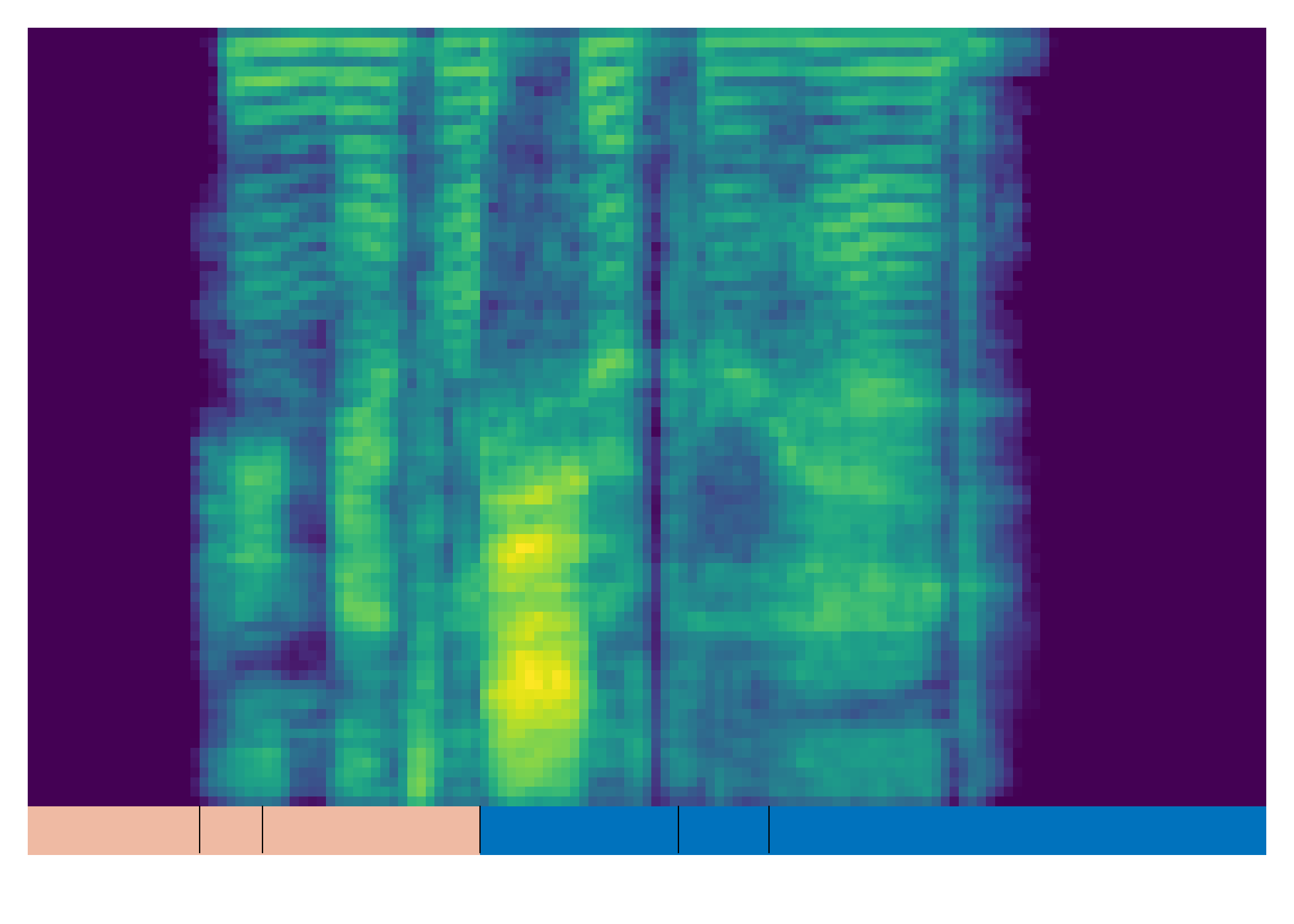}
	\caption{Augmentation}
    \end{subfigure}
\caption{An example data augmentation demonstrating a non-terminal VP $\to$ VP constituency substitution.}
\vspace{-0.3cm}
\label{fig:sub}
\end{figure}

Tree Substitution Grammar \cite{Schabes1990MathematicalAC} formalizes language generation as rewriting of non-terminal nodes of a tree (a parsed representation of a sentence), each with some arbitrary size subtrees. Given a text corpus with parses (e.g. constituency parse), these formulations allow estimating a generative  grammar of the language \cite{DBLP:journals/jmlr/CohnBG10}.
Equivalently, without estimating the actual grammar, this formulation allows creation of new text samples by substituting a subtree with another subtree from the corpus (given that both the subtrees have the same type of non-terminal node). This is a popular data augmentation technique which not only works for low-resource text corpora but also shows gains for augmentation with massively pretrained models \cite{subsub2021}. An example of creating augmented text by substituting a constituency parse subtree can be found in \Cref{fig:sub}.

The following describes how we apply the subtree substitution technique for augmenting a TTS dataset.
Assume we have a dataset $D$ consisting of triples $e_i = \langle t_i, x_i, a_i \rangle$, where $t_i$ denotes text features, $x_i$ denotes audio features and $a_i$ denotes the alignment between text and audio.
For each pair of training examples $e_i$, $e_j$ we take their respective texts ($t_i$, $t_j$) and parse them using a constituency parser.
Then, for each constituent $c^i_m$ in the parse tree of $t_i$ and each constituent $c^j_n$ in the parse tree of $t_j$, we create an augmented text by substituting $c^i_m$ by $c^j_n$, provided that $c^i_m$ and $c^j_n$ have the same constituency type.
This procedure allows to create multiple new texts for each pair $e_i$, $e_j$.
Then, we create augmented audio features taking advantage of available alignments $a_i$ and $a_j$.
We simply concatenate appropriate fragments of audio features, following the alignments.

\begin{figure}[!ht]
\includegraphics[scale=0.45]{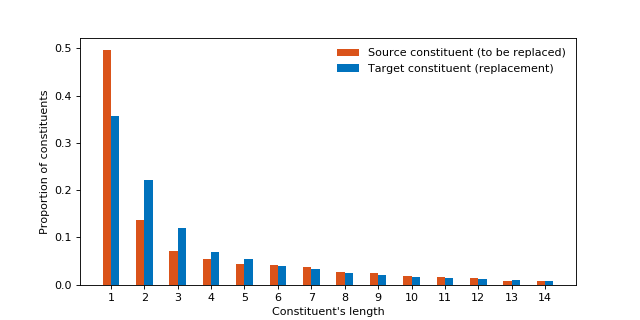}
\caption{\label{const_len_dist} Distributions of constituent lengths used to create augmented dataset.}
\vspace{-0.3cm}
\end{figure}

Note that this procedure only allows to create an augmented example from exactly two original examples.
Also, for each augmented example there is exactly one subtree substitution.
Thus, audio joints (as described in \Cref{sec:perm}) are quite sparse: there is no more than two per utterance.
\Cref{const_len_dist} presents distributions of lengths (in number of words) of parse tree constituents that were used to augment dataset $D_1$ (see \Cref{sec:datasetup}).
As can be seen, substituted constituents are usually short (1-3 words long).
On the other hand, the remaining part of the utterance that surrounds substituted constituent is left intact and is usually much longer.
This ensures that long enough audio pieces from original samples are presented to the model during training so it can learn consistent prosody spanning over longer fragments of text.

\section{Experiments}
\label{sec:experiments}

\subsection{Model architectures}
\label{ssec:subhead}

\begin{figure}[!ht]
\includegraphics[scale=0.9]{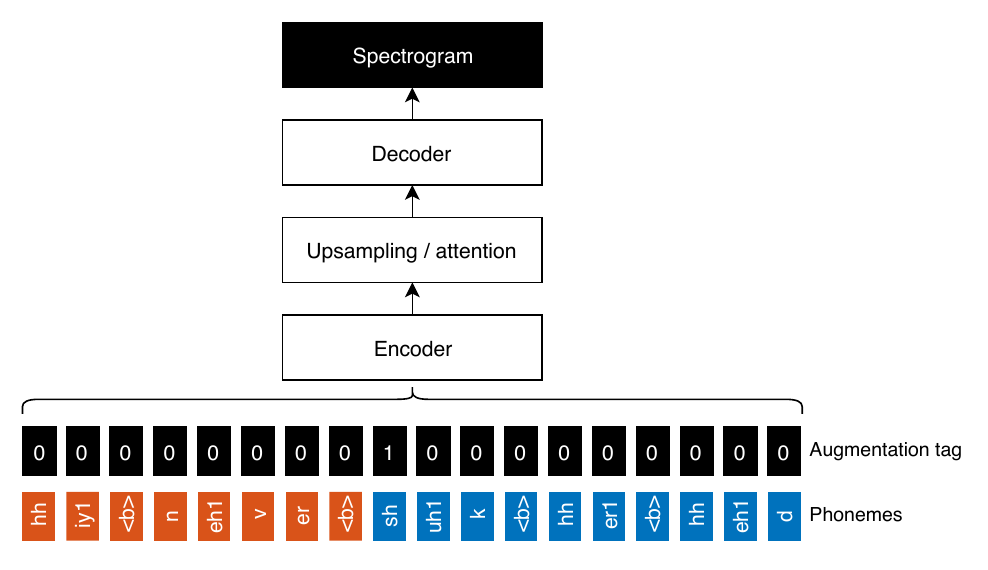}
\caption{\label{architecture} High-level architecture of models used in experiments. We use two model variants: one with attention between phoneme encodings and frames and another with attention replaced by upsampling. Input phonemes are concatenated with per-phoneme augmentation conditionings.}
\vspace{-0.4cm}
\end{figure}

We apply our proposed method to two types of architectures (\Cref{architecture}): an attention-based architecture (A) and an architecture using externally provided durations (B). We choose them because they are the two main classes of models currently used in TTS for modelling mel-spectrograms. The attention-based one largely follows the architecture of Tacotron 2 \cite{taco2}. The phoneme encoder consists of a stack of 1D convolutions and a bi-LSTM and is followed by a location-sensitive attention mechanism \cite{chorowski2015attention}. Finally, the autoregressive decoder predicts the mel-spectrogram frames.

The system using externally provided durations is inspired by non-attentive Tacotron \cite{shen2020non}, although we introduce some simplifications. Our variant consists of two main components: a TTS model and a separate duration model. The TTS model has the same architecture as the attention-based model described before, the only difference between the systems being that the attention mechanism is replaced by upsampling from phoneme to frame level using oracle or externally predicted durations. The upsampling is followed by a bi-LSTM layer that replaces Gaussian Upsampling from \cite{shen2020non}. The separately trained external duration model has the same architecture as the phoneme encoder with an additional dense layer and is trained on the oracle durations using an L2 loss.

The only addition to the models that is specific to the proposed method is that we introduce a phoneme-level conditioning consisting of a binary tag indicating whether the phoneme is the first phoneme after a boundary between two utterances that were joined together.
For both types of model, our frontend converts the input text into phonemes, which are input to the phoneme encoder. We use a multispeaker parallel WaveNet as a vocoder to produce the waveform for evaluations \cite{uv2021}.

\subsection{Data preparation}
\label{sec:datasetup}

We run our experiments on two datasets:

\begin{itemize}
    \item A proprietary dataset containing highly expressive English speech recorded by a female voice talent ($D_1$).

    \item Two high-quality English voices selected from public HiFi TTS dataset \cite{bakhturina2021hi}, one male ($D_2^m$) and one female ($D_2^f$).

\end{itemize}

We partition $D_1$ and $D_2$ into training, evaluation and test sets.
To evaluate the proposed method in different data settings, we randomly select three subsets of sizes 10h, 5h and 2h from the $D_1$ train set.
In the experiments with datasets $D_2^m$ and $D_2^f$ we always use 10h of data randomly selected from the training set.
To facilitate trainings, we extract mel-spectrograms with 80 mel channels, using 12.5 ms frame shift and 50 ms frame length.
We use proprietary TTS frontend to extract phonetic transcriptions of the audios and the Kaldi toolkit \cite{povey2011kaldi} to find alignments between text and audio.

In order to generate augmented dataset we apply the procedure described in \Cref{sec:parse}. 
For constituency parsing we use Berkeley Neural Parser \cite{kitaev2018constituency}. 
Note that it allows to produce millions of new samples from few thousand of original utterances. 
To simplify our experiments, we randomly select 500k augmented examples for each dataset.
This roughly corresponds to 1k hours of speech. 
However, the augmented data does not offer more acoustic diversity than the original dataset. 
This is why we always report results for the number of hours of the original dataset.

\subsection{Evaluation setup}
\label{sec:expsetup}

We evaluate our models using the following protocol. For each tested configuration we randomly select 100 utterances from the testset of the respective dataset. We synthesize speech samples for these utterances using the baseline and the proposed model. Then we ask 60 native English speakers to rate each pair of samples in a preference test (we ask them to choose better sounding version).

For the attention-based variant we additionally run a robustness test measuring the word error (WER) rate and the phoneme error rate (PER) on the baseline and proposed models. This is because such models, trained on highly expressive data, are particularly prone to intelligibility issues such as mispronunciations, mumbling, skipping or repeating phonemes, or cut-offs. We run the robustness tests on a text corpus containing 2500 diverse utterances.

\subsection{Results}
Here we refer to the models developed without using any augmented data as \textit{baseline model} (or simply \textit{baseline}) and the models developed with augmented data as \textit{proposed models}.

\subsubsection{Attention-based models (A)}

For experiments with architecture (A) we use 5h and 10h variants of $D_1$.
For each variant of $D_1$ we train one baseline model and one proposed model.
We train the models for 200k steps using SGD with Adam optimizer.
We also trained 2h variants, but the baseline failed to produce intelligible speech in this setting, and thus we do not evaluate it. At the same time the proposed model produces intelligible speech with only 2 hours of training data, which suggests that the impact of our method increases in lower data settings.

In order to demonstrate that the proposed method reduces overfitting, we calculate test set loss for both the baseline and proposed model. As can be seen in \Cref{tab:preferenceA}, the proposed models achieve lower test set losses than respective baselines. This shows that the proposed method indeed reduces overfitting.

Then, we evaluate the robustness of the trained models by running an ASR model on synthesized audio and comparing it to the text. The WER and PER are shown in \Cref{tab:stability}. The proposed method brings a considerable improvement in terms of WER and PER when training on 5 or 10 hours of data.
The higher error rates for the proposed system trained on 10h compared to 5h is a counterintuitive result.
However, given the relatively small difference compared to the differences between the baseline and proposed models, we argue that this can be attributed to the stochasticity of training the models, selecting the data, and running the robustness analyses.

Finally, we run preference tests to compare the baseline and proposed models. \Cref{tab:preferenceA} summarizes the results of the preference tests that render the proposed models being statistically significantly preferred over baseline (with p-value $< 0.05$). As the robustness test results suggest, the baselines suffer from robustness issues, so for the evaluation we filter out samples with nonzero WER, moving the results even more in favor of our model.

\begin{table}
\small\centering
\begin{tabular}{l|rr|rrr}
\toprule
    \multirow{2}{*}{Dataset} & \multicolumn{2}{c|}{WER} & \multicolumn{2}{c}{PER} \\
\cmidrule{2-5}
     & Base  & Ours  & Base & Ours  \\
    \midrule
    $D_1^{\mathrm{10h}}$  & 0.74 &  \textbf{0.26} &  0.59 & \textbf{0.07}  \\
    $D_1^{\mathrm{5h}}$  & 1.00 &  \textbf{0.19} &  1.00 & \textbf{0.05}  \\
\bottomrule
\end{tabular}
    \caption{Word and phoneme error rates (bolded is better) of attention-based models for baseline (Base) and proposed models (Ours), relative to the WER and PER of 5h baseline model.}
\vspace{-0.1cm}
\label{tab:stability}
\end{table}

\begin{table}
\small\centering
\begin{tabular}{l|rrr|rr}
\toprule
    \multirow{2}{*}{Dataset} & \multicolumn{3}{c|}{\% Preference} & \multicolumn{2}{c}{Test loss} \\
\cmidrule{2-6}
     & Base  & Ours  &  None & Base & Ours  \\
\midrule
    $D_1^{\mathrm{10h}}$  & 38 &  \textbf{41.1} &  20.9 & 0.034 & 0.028  \\
    $D_1^{\mathrm{5h}}$  & 33.5 &  \textbf{46} &  20.5 & 0.039 & 0.031 \\

\bottomrule
\end{tabular}
    \caption{Preference ratings (bolded is preferred) between baselines (Base) and proposed models (Ours) with architecture (A).}
\vspace{-0.5cm}
\label{tab:preferenceA}
\end{table}

\vspace{-0.1cm}
\subsubsection{Models using externally provided duration (B)}

In order to evaluate the proposed method on architecture (B) we train baseline and proposed models on datasets $D_1$, $D_2^m$ and $D_2^f$. For dataset $D_1$ we train 10h, 5h and 2h variants. All TTS models are trained for 100k steps using SGD with the Adam optimizer.

\Cref{tab:preferenceB} presents the average loss calculated for each model on the respective test set. For all datasets, the proposed models achieve a lower test set loss, which demonstrates that our proposed method reduces overfitting of the TTS model. \Cref{tab:preferenceB} demonstrates that the proposed models are preferred by listeners. All results are statistically significant with p-value $< 0.05$. These results show that our method not only reduces overfitting for architecture (B) but also improves quality of speech produced by the model. Note that there is a bigger difference in preference for the lowest resource setting (2h) than in the other ones. This is another data point suggesting that the effectiveness of our method increases with lower amount of data.

\begin{table}
\small\centering
\begin{tabular}{l|rrr|rr}
\toprule
    \multirow{2}{*}{Dataset} & \multicolumn{3}{c|}{\% Preference} & \multicolumn{2}{c}{Test loss} \\
\cmidrule{2-6}
& Base  & Ours  &  None & Base & Ours   \\
\midrule
    $D_1^{\mathrm{10h}}$  & 35.6 &  \textbf{40.3} &  24.1  & 0.035 & 0.034 \\
    $D_1^{\mathrm{5h}}$  & 37 &  \textbf{41.4} &  21.6 & 0.037 & 0.036  \\
    $D_1^{\mathrm{2h}}$  & 34.7 &  \textbf{43.2} &  22.1 & 0.040 & 0.039  \\
    $D_2^m$  & 34.5 &  \textbf{38.1} &  27.4  & 0.047  & 0.045  \\
    $D_2^f$  & 35.3 &  \textbf{38.3} &  26.4  & 0.048 & 0.047 \\
\bottomrule
\end{tabular}
    \caption{Preference ratings (bolded is preferred) between baselines (Base) and proposed models (Ours) with architecture (B).}
\vspace{-0.1cm}
\label{tab:preferenceB}
\end{table}

\vspace{-0.1cm}
\subsubsection{Ablation study: Importance of augmentation conditioning}

In this section we investigate the impact of the extra augmentation conditioning.
To this end, we train a model with architecture (B) on the 10 hours variant of $D_1$ dataset augmented using our method, but without applying the additional conditioning.
Without conditioning, the augmentation no longer brings improvement over the baseline in the tested setting. \Cref{tab:cond_base} presents the results of a preference test between the proposed model without conditioning and the baseline model.
The baseline has higher preference ratings, although the result is not statistically significant (p-value $> 0.05$).
We also confirm that the proposed model with conditioning is significantly preferred over the model without conditioning (see \Cref{tab:cond_base} for detailed results that are significant with p-value $< 0.05$).
This suggests that using the augmentation conditioning is important for our method.

\begin{table}
\small\centering
\begin{tabular}{l|r>{\raggedleft}p{1.6cm}>{\raggedleft}p{1.6cm}r}
\toprule
\multirow{2}{*}{Dataset} & \multicolumn{4}{c}{\% Preference} \\
\cmidrule{2-5}
 & Base  & W/o conditioning  &  With conditioning &  None  \\
\midrule
    \multirow{2}{*}{$D_1^{\mathrm{10h}}$}  & 39.8 &  38.3 & - & 21.9  \\
     & -  & 37.9 &  \textbf{41.7} &  20.4  \\
\bottomrule
\end{tabular}
\caption{Preference ratings (bolded is preferred) for: 1) baseline (Base) vs proposed model without augmentation conditioning (W/o conditioning); 2) proposed model without augmentation conditioning (W/o conditioning) vs with conditioning (With conditioning).}
\vspace{-0.5cm}
\label{tab:cond_base}
\end{table}

\vspace{-0.1cm}
\section{Conclusion}
\label{sec:conclusion}

In this paper, we present a novel data augmentation technique for TTS that can considerably increase the diversity of text conditionings seen in training. We create new training samples by parse tree constituents substitution in both text and audio. We take inspiration from existing work on data augmentation for generative modelling to address the problem of augmented samples being out-of-distribution.

We show that our technique reduces overfitting in TTS models trained on expressive data. This overfitting reduction translates into improved speech quality according to perceptual tests. We showed effectiveness of our technique on a range of voices and architectures. Additionally, for attention-based architectures we demonstrated that our method greatly improves model's robustness.

\bibliographystyle{IEEEbib}
\bibliography{refs}

\end{document}